\documentclass[10pt]{amsart}

\theoremstyle{plain}
\newtheorem{lemma}{Lemma}[section]
\newtheorem{definition}[lemma]{Definition}
\newtheorem{corollary}[lemma]{Corollary}

\newtheorem{theorem}[lemma]{Theorem}

\theoremstyle{remark}

\newcommand{\n}{\mathbf{n}}
\newcommand{\0}{\mathbf{0}}
\newcommand{\N}{\mathbb{N}}

\renewcommand{\c}{\mathrm{c}}

\newcommand{\mylabel}[1]{\label{#1}}

\newcommand{\G}{\mathcal{G}}

\newcommand{\concat}{\mathit{concat}}
\newcommand{\lconcat}{\mathit{lconcat}}
\newcommand{\rconcat}{\mathit{rconcat}}
\newcommand{\linsert}{\mathit{linsert}}
\newcommand{\rinsert}{\mathit{rinsert}}
\newcommand{\lwrap}{\mathit{lwrap}}
\newcommand{\rwrap}{\mathit{rwrap}}
\newcommand{\disconn}{\mathit{disconn}}
\newcommand{\lembed}{\mathit{lembed}}
\newcommand{\rembed}{\mathit{rembed}}
\newcommand{\lp}{\mathit{loop}}
\newcommand{\stack}{\mathit{nest}}
\newcommand{\inter}{\mathit{cross}}

\newcommand{\astack}[2]{\left(\!\!\!\begin{array}{c}#1\\#2\end{array}\!\!\!\right)}
\newcommand{\pstack}[6]{\left(\!\!\!\begin{array}{c@{\,,\,}c}#1&#3\\#4&#6\end{array}\!\!\!\right)}
\newcommand{\blank}{\circ}

\newcommand{\setmins}{\!\setminus\!}

\newcommand{\id}{\mathit{id}}
\newcommand{\empt}{\mathit{empty}}

\usepackage{amssymb}
\usepackage{amsthm}
\usepackage[algoruled,titlenotnumbered]{algorithm2e}

\usepackage{xy}
\xyoption{all}
\xyoption{poly}

\newcommand{\RNA}{\mathrm{RNA}}

\begin{document}

%
%

\title[Structural Alignments of pseudo-knotted RNA-molecules]{Structural Alignments of pseudo-knotted RNA-molecules in polynomial
  time}

\author{Michael Brinkmeier}

\begin{abstract}
  An RNA molecule is structured on several layers. The primary and most
  obvious structure is its sequence of bases, i.e. a word over the alphabet
  $\{A,C,G,U\}$. The higher structure is a set of one-to-one base-pairings
  resulting in a two-dimensional folding of the one-dimensional sequence. One
  speaks of a secondary structure if these pairings do not cross and of a
  tertiary structure otherwise.

  Since the folding of the molecule is important for its function, the search
  for related RNA molecules should not only be restricted to the primary
  structure. It seems sensible to incorporate the higher structures in the
  search. Based on this assumption and certain edit-operations a distance
  between two arbitrary structures can be defined. It is known that the
  general calculation of this measure is NP-complete \cite{zhang02similarity}.
  But for some special cases polynomial algorithms are known. Using a new
  formal description of secondary and tertiary structures, we extend the class
  of structures for which the distance can be calculated in polynomial time.
  In addition the presented algorithm may be used to approximate the
  edit-distance between two arbitrary structures with a constant ratio.
\end{abstract}

\maketitle

\section{Introduction}

Ribonucleic acid (RNA) is structured on three levels. The primary and most
obvious structure is the underlying sequence of bases. The higher layers of
structure are given by its folding, i.e. its pattern of base pairings. As long
as the structure is {\em nested}, one speaks of a secondary structure. If {\em
  crossed} pairs or {\em pseudoknots} exist, the molecule is of tertiary
structure\footnote{There exist two definitions of tertiary structure, 
one as given above, and an alternative definition as the spatial arrangement
of bases.}. Since the folding and the embedding into the
three-dimensional space are important for the functional properties of an RNA
molecule, it is of some interest to compare different molecules based on the
secondary and tertiary structure and not only on the primary structure.

Restricting to the primary structure, the comparison of two or more RNA
strands is efficiently solvable by the same techniques used for the alignment
of DNA sequences \cite{gusfield97book}. For a given set of (weighted) edit
operations, the {\em edit-distance}, i.e. the minimal number of operations
needed for the transformation of one sequence into the other, is calculated. 
This results in an {\em alignment} of the two structures.

In the literature this approach is transferred to higher structures of RNA
mole\-cules in various ways. Some of them rely on the tree representation of
secondary structures and measure the {\em tree-edit-distance}
(eg. \cite{zhang90comparing, zhang96constrained, zhang96efficient}), some involve {\em stochastic context-free 
  grammars} \cite{sakakibara94recent, sakakibara94stochastic}. Additional methods can be found in
\cite{bafna95computing} and \cite{lenhof98polyhedral}. But most of them,
mostly due to formal restrictions, cover only nested sequences.

As K. Zhang et al. do in \cite{zhang02similarity} and
\cite{zhang02editdistance}, we treat unpaired bases and base-pairings as
atomic units. There exist several variations of this approach. First of
all, one has to choose a set of basic edit-operations. Secondly, one has to
chose the scores or weights for these operations, which may possibly depend on
the underlying bases. Since we view unpaired bases and pairings as unbreakable
units, it is only allowed to replace an unpaired base/pairing by another one,
to remove it or to introduce a new one. It is not allowed to remove only 
one partner of a pairing, to break a pairing without removing the bases etc.

These edit-operations are quite restrictive, compared to the more flexible set
used in \cite{zhang02editdistance}. But nonetheless, the general problem of
finding a minimal alignment (sequence of edit-operations) is NP-complete, as
long as the edit-operations do rely on the higher structures, i.e. the
pairings (this is known for several choices of edit-operations
(\cite{zhang02similarity} \cite{zhang02editdistance}). So far 
polynomial algorithms only are known for alignments of a secondary
structure with an arbitrary one\footnote{The algorithm of Zhang can be
  adapted to cover certain restricted H-like pseudoknots instead of nested  
  structures. But the details are not given in the cited paper}.

We are going to present a new description of higher structures of RNA , using a
formal system closely related to graph grammars. In the first section the 
formalism is defined and a set of generators, i.e. certain small structures
which are used to construct larger, so-called {\em decomposable} structures,
is presented. As we will see, this description includes nested structures and
a wide variety of pseudoknots, but not all. Following that, we are going to
describe our edit-operations and the resulting type of alignment of two
tertiary structures. Connecting this with our formalism leads to the essential
observation that the alignments of decomposable structures are of a special
type, called {\em semi-decomposable}, meaning that a core structure of the
alignment, consisting of all matched/mismatched pairings and bases, still is
decomposable. This results in the framework of an algorithm calculating the
{\em exact} score of a minimal alignment of a so-called {\em decomposable}
structure with an arbitrary one. This algorithm runs in a time polynomial in
the number of bases of the aligned sequences and polynomial space. The degrees 
of the polynomials depend on the choice of generators. In fact it is possible
to extend the set of decomposable strucutres easily by the introduction of additional 
generators. This increases the degree of the polynomials, but the required time and 
space remain polynomial, even though the general problem is NP-complete.
 
Finally we will give some results about the approximation ratio of the algorithm for 
arbitrary pairs of RNA molecules, depending on the chosen scores.

\begin{figure}
\begin{center}
\begin{tabular}{|c|c|c|c|c|}
  \hline
  {\bf Covered structures} & {\bf time} & {\bf space} & \\
  \hline
  (plain, plain) & $O(n^2)$ & $O(n^2)$ &  eg. \cite{kruskal83timewraps} \\ \hline
  (nested, any) & $O(n^4)$ & $O(n^2)$ &  \cite{zhang02similarity} \\ \hline
  (decomposable, any) & $O(n^{12})$ &  $O(n^8)$ & this paper \\ \hline
  (any,any) & \multicolumn{2}{c|}{NP-complete} & \cite{zhang02similarity}  \\ \hline
\end{tabular}
\end{center}
\caption{Complexities of alignment algorithms}
\end{figure}

Unfortunately the runtime and especially the space requirements prohibit an
actual implementation of this algorithm. But nonetheless, they may be the
foundation of a family of more efficient algorithms. Ideas, how this basic
algorithm can be improved, are given in the last section of this paper.


\section{Higher RNA Structures}


Usually an RNA molecule is represented by a sequence of bases, i.e. a word $s$
over the alphabet $\Sigma_{\RNA} = \{ A,C,G,U\}$ together with a set of {\em 
pairings}, i.e. a set of pairs $(i,j)$ with $1 \leq i < j \leq |s|$. Graphically
these are represented by a structure (multi-)graph with vertex set $\{1, \dots ,|s|\}$. 
The $i$-th base corresponds to the vertex $i$, and two subsequent vertices/bases are 
connected by a {\em backbone}(-edge). In addition each pairing $(i,j)$ is represented 
by an edge between $i$ and $j$.

\begin{figure}[b]
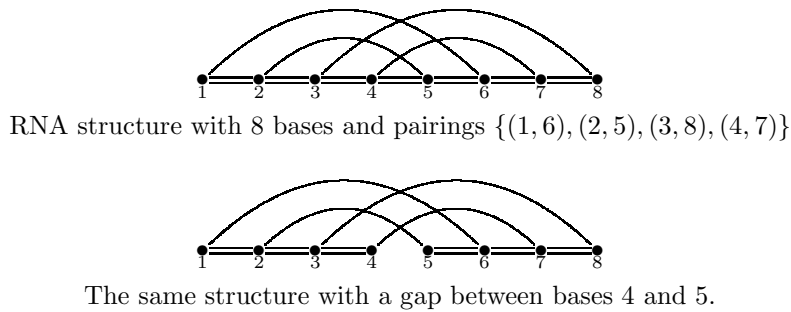

\begin{center}
  $\xy
  <0.75cm,0cm>:
  (0,0)*+={\bullet}="0",(0,-0.2)*={\scriptstyle{1}},
  (1,0)*+={\bullet}="1",(1,-0.2)*={\scriptstyle{2}},
  (2,0)*+={\bullet}="2",(2,-0.2)*={\scriptstyle{3}},
  (3,0)*+={\bullet}="3",(3,-0.2)*={\scriptstyle{4}},
  (4,0)*+={\bullet}="4",(4,-0.2)*={\scriptstyle{5}},
  (5,0)*+={\bullet}="5",(5,-0.2)*={\scriptstyle{6}},
  (6,0)*+={\bullet}="6",(6,-0.2)*={\scriptstyle{7}},
  (7,0)*+={\bullet}="7",(7,-0.2)*={\scriptstyle{8}}

  \ar@{=}"0";"1"
  \ar@{=}"1";"2"
  \ar@{=}"2";"3"
  \ar@{=}"3";"4"
  \ar@{=}"4";"5"
  \ar@{=}"5";"6"
  \ar@{=}"6";"7"
  \ar@{-}@`{(2.5,2.5)}"0";"5"
  \ar@{-}@`{(2.5,1.5)}"1";"4"
  \ar@{-}@`{(4.5,2.5)}"2";"7"
  \ar@{-}@`{(4.5,1.5)}"3";"6"
  \endxy $
 
  \vspace*{2mm}
  
  RNA structure with 8 bases and pairings $\left\{ (1,6),(2,5),(3,8),(4,7)\right\}$
  
  \vspace*{5mm}

  $\xy
  <0.75cm,0cm>:
  (0,0)*+={\bullet}="0",(0,-0.2)*={\scriptstyle{1}},
  (1,0)*+={\bullet}="1",(1,-0.2)*={\scriptstyle{2}},
  (2,0)*+={\bullet}="2",(2,-0.2)*={\scriptstyle{3}},
  (3,0)*+={\bullet}="3",(3,-0.2)*={\scriptstyle{4}},
  (4,0)*+={\bullet}="4",(4,-0.2)*={\scriptstyle{5}},
  (5,0)*+={\bullet}="5",(5,-0.2)*={\scriptstyle{6}},
  (6,0)*+={\bullet}="6",(6,-0.2)*={\scriptstyle{7}},
  (7,0)*+={\bullet}="7",(7,-0.2)*={\scriptstyle{8}}

  \ar@{=}"0";"1"
  \ar@{=}"1";"2"
  \ar@{=}"2";"3"
  \ar@{=}"4";"5"
  \ar@{=}"5";"6"
  \ar@{=}"6";"7"
  \ar@{-}@`{(2.5,2.5)}"0";"5"
  \ar@{-}@`{(2.5,1.5)}"1";"4"
  \ar@{-}@`{(4.5,2.5)}"2";"7"
  \ar@{-}@`{(4.5,1.5)}"3";"6"
  \endxy $

  \vspace*{2mm}
  
  The same structure with a gap between bases 4 and 5.
     
\end{center}
  \caption{\label{figexample1}Examples of RNA structures}
\end{figure}

We require an additional type of RNA-structures, introducing a gap between two bases,
as already used by Rivas and Eddy in \cite{rivas99dynamic}. These will be called 
{\em gapped} or {\em 1-structures} (An example is given in Fig. \ref{figexample1}).
In addition we are going to seperate the structure from the sequence of bases, providing
us with a simplified notation.


For $n\in\N$ and $n\geq 1$ let $\n$ denote the finite set $\{1,\dots, n\}
\subset \N$ and $\0$ the empty set.

\begin{definition}[{{\bf 0- and 1-Structures}}]
  \mylabel{defstructure}
  A {\em 0-structure} $\sigma = (n,P)$ (or {\em structure of type $0$})
  consists of a natural number $n$ and a set $P\subset\n\times\n$ of pairs
  $(i,j)$ with $i<j$, such that for $(i,j), (i',j') \in P$ the intersection
  $\{i,j\}\cap \{i',j'\}$ is either empty or $\{i,j\} = \{i',j'\}$.
  
  A {\em 1-Structure} $\sigma =(n,P,k)$ (or {\em structure of type $1$})
  consists of a $0$-structure $(n,P)$ and a natural number $0 \leq k \leq n$.
\end{definition}

The elements of $\n$ are the {\em bases} and the pairs $(i,j) \in P$ the
{\em (base-)pairings}. A base $i \in \n$ of $\sigma$ is {\em paired} if
there exist a pairing $(i,j)$ or $(j,i)$, and {\em unpaired}
otherwise. Furthermore each base is paired with at most one other base.
The pairings and unpaired bases are also called {\em structural elements}. 
In a 1-structure $\sigma=(n,P,k)$ the sequence of bases is split after 
base $k$ into two intervals $[i,k]$ and $[k+1,n]$ called {\em legs}.

We explicitly allow the legs to be empty (by setting $k=0$ and $k=n$). This
allows us to view 0-structures as 1-structures with one empty leg. Therefore 
we may use these cases to relate to 0- and 1-structures without mentioning them
explicitly.

Since all pairings $(i,j)$ satisfy $i<j$, we do not need to differentiate
between $(i,j)$ and $(j,i)$. Furthermore, if a pairing $(i,j)$ is given
we assume $i < j$, unless stated otherwise. Furthermore we simply write
$(i,j)\in\sigma=(n,P)$ for pairings in $P$ and $i\in\sigma$ for unpaired
bases.

\begin{figure}
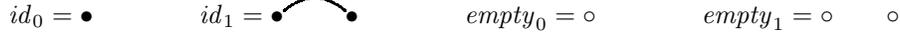

\begin{center}
$
\id_0 = \bullet \qquad\qquad
\id_1 =  \xy  <1cm,0cm>:
  (0,0)*+={\bullet}="0",
  (1,0)*+={\bullet}="1"
  \ar@{-}@`{(0.5,0.5)}"0";"1"
  \endxy \qquad\qquad
\empt_0 = \circ \qquad\qquad
\empt_1 = \circ\qquad\circ
$
\end{center}
\caption{\label{figtrivial}The trivial structures}
\end{figure}

There exist four {\em trivial} structures containing at most one structural
element. The {\em identities} simply consist of a free base or a single
pairing. The {\em empty structures} are structures with no bases. In the
graphical representation  (see Fig. \ref{figtrivial}) we use an empty circle
``$\circ$'' for empty legs.

\begin{figure}[t]
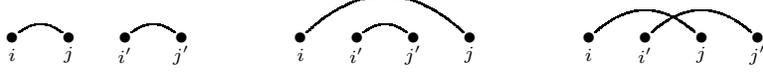

\begin{center}
$
  \xy
  <0.75cm,0cm>:
  (0,0)*+={\bullet}="0",(0,-0.3)*={\scriptstyle{i}},
  (1,0)*+={\bullet}="1",(1,-0.3)*={\scriptstyle{j}},
  (2,0)*+={\bullet}="2",(2,-0.3)*={\scriptstyle{i'}},
  (3,0)*+={\bullet}="3",(3,-0.3)*={\scriptstyle{j'}}
  \ar@{-}@`{(0.5,0.5)}"0";"1"
  \ar@{-}@`{(2.5,0.5)}"2";"3"
  \endxy
  \qquad\qquad
  \xy
  <0.75cm,0cm>:
  (0,0)*+={\bullet}="0",(0,-0.3)*={\scriptstyle{i}},
  (1,0)*+={\bullet}="1",(1,-0.3)*={\scriptstyle{i'}},
  (2,0)*+={\bullet}="2",(2,-0.3)*={\scriptstyle{j'}},
  (3,0)*+={\bullet}="3",(3,-0.3)*={\scriptstyle{j}}
  \ar@{-}@`{(1.5,1.5)}"0";"3"
  \ar@{-}@`{(1.5,0.5)}"1";"2"
  \endxy
  \qquad\qquad
  \xy
  <0.75cm,0cm>:
  (0,0)*+={\bullet}="0",(0,-0.3)*={\scriptstyle{i}},
  (1,0)*+={\bullet}="1",(1,-0.3)*={\scriptstyle{i'}},
  (2,0)*+={\bullet}="2",(2,-0.3)*={\scriptstyle{j}},
  (3,0)*+={\bullet}="3",(3,-0.3)*={\scriptstyle{j'}}
  \ar@{-}@`{(1,1)}"0";"2"
  \ar@{-}@`{(2,1)}"1";"3"
  \endxy
$
\end{center}
\caption{\label{figpairings}Independent, nested and crossing pairings}
\end{figure}

Two pairings $(i,j)$ and $(i',j')$ with $i<i'$ are called {\em independent}
if $i<j<i'<j'$, {\em nested} if $i<i'<j'<j$ and {\em crossed} if $i<i'<j<j'$.
A 0- or 1-structure $\sigma$ is called {\em nested} if any two pairings
are either independent or nested (see Fig. \ref{figpairings}). A structure 
containing crossed pairings is called {\em pseudoknot} or {\em pseudoknotted}.

The natural numbers induce a total order on the structural elements. More 
precisely we define
\begin{definition}[{{\bf The Order of Structural Elements}}]
  Let $\sigma$ be a structure. The relation $<_{\sigma}$ on
  the set of structural elements is given by:
  \begin{itemize}
  \item $i <_{\sigma} j$ if and only if $i < j$.
  \item $i <_{\sigma} (i',j')$ if and only if $i < i'$.
  \item $(i',j') <_{\sigma} i$ if and only if $i' < i$. 
  \item $(i,j) <_{\sigma} (i',j')$ if and only if $i < i'$. 
  \end{itemize}
\end{definition}

\subsection{Structures and Sequences}

Up to this point we only described the structure of RNA-molecules, but not the 
sequence of bases.  In fact we put some effort in the seperation of
the secondary structures from the actual sequence of nucleotides. 

%

\begin{definition}[{{\bf Folded $\Sigma$-Sequences}}]
  \mylabel{defsequence}
  Let $\Sigma$ be an alphabet, i.e. a finite non-empty
  set. A {\em $\Sigma$-sequence of type 0} is a word in $\Sigma^{\ast}$ and a
  {\em $\Sigma$-sequence of type 1} is a pair $s=(s_1,s_2)$ of words in
  $\Sigma^{\ast}$.

  A {\em folded $\Sigma$-sequence} is a pair $(\sigma,s)$
  consisting of a structure $\sigma$ and a $\Sigma$-sequence $s$ of the same
  type as $\sigma = (n,P,k)$, such that 
  \begin{itemize}
  \item $|s| = n$ if $\sigma$ is a $0$-structure and
  \item $|s_1| = k$ and $|s_2| = n-k$ if it is a $1$-structure.
  \end{itemize}

\end{definition}

For RNA molecules the standard alphabet is $\Sigma_{\RNA} = \{ A,C,G,U \}$. 
But nonetheless it is useful to use this general definition, because in the context of 
alignmets bases may be deleted or inserted, resulting in ``empty'' bases, which correspond
to the fifth letter ``$\circ$''.

%
%
%

\subsection{The Compositions}

We will especially consider structures constructed from smaller ones by two basic operations,
called {\em compositions}. The first replaces an unpaired base of an arbitrary structure by
a whole $0$-structure. The second composition does exaclty the same with a pairing and a 
1-structure. More formally we use the following definition.


\begin{definition}[{{\bf The Compositions}}] \hfill

\begin{enumerate}
  \item Let $\sigma = (n,P,k)$ be a structure and $i$ one of its unpaired
  bases. For each 0-structure $\tau = (m,Q)$ the structure
  $\sigma\circ_i\tau$ is defined by
  $$ \sigma\circ_i\tau := (n+m-1,P\circ_i Q,k') $$
  with 
  $$ k' := \begin{cases}
    k & \text{ if } k < i \\
    k+m-1 & \text{ if } k \geq i
  \end{cases} $$
  and $P\circ_i Q$ contains the following pairings
  \begin{itemize}
  \item $(j_1',j_2')$ for all $(j_1,j_2) \in P$ with 
    $ j_x' = \begin{cases}
    j_x & \text{ if } j_x < i \\
    j_x + m - 1 & \text{ if } i < j_x
    \end{cases} $
  \item $(j_1+i,j_2+i)$ for all $(j_1,j_2) \in Q$.
  \end{itemize}
  The operation $-\circ_i-$ is called {\em composition along the (unpaired)
  base $i$}.
  \vspace*{3mm}
\item
  Let $\sigma = (n,P,k)$ be a structure and $(i,j)$ one of its
  pairings. For each 1-structure $\tau = (m,Q,l)$ the structure
  $\sigma\circ_{(i,j)}\tau$ is defined by
  $$ \sigma\circ_{(i,j)}\tau := (n+m-2,P\circ_{(i,j)} Q,k') $$
  with 
  $$ k' := \begin{cases}
    k & \text{ if } k < i \\
    k+l-1  & \text{ if } i \leq k < j \\
    k+m-2 & \text{ if } j \leq k
  \end{cases} $$
  and $P\circ_{(i,j)} Q$ contains the following pairings
  \begin{itemize}
  \item $(j_1',j_2')$ for all $(j_1,j_2) \in P\setminus\{(i,j)\}$ with 
    $ j_x' = \begin{cases}
    j_x & \text{ if } j_x < i \\
    j_x + l -1 & \text{ if } i < j_x < j \\
    j_x + m -2 & \text{ if } j < j_x
    \end{cases}$
  \item $(j_1',j_2')$ for all $(j_1,j_2) \in Q$ with
    $ j_x' = \begin{cases}
       j_x + i -1 & \text{ if } j_x < l \\
       j_x + j -1 & \text{ if } l \leq j_x.
    \end{cases}$
  \end{itemize}
  The operation $-\circ_{(i,j)}-$ is called {\em composition along the pairing $(i,j)$}.
\end{enumerate}
\end{definition}

\begin{figure}
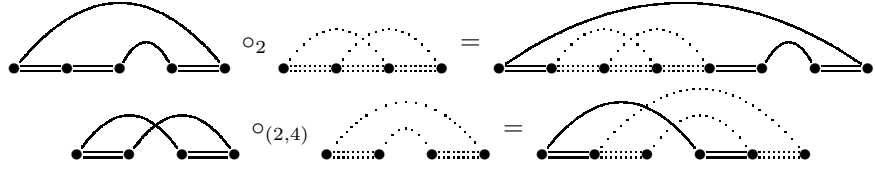

  \begin{center}
  $$ \xy <0.7cm,0cm>:
  (0,-0.5)*+={\bullet}="0",
  (1,-0.5)*+={\bullet}="1",
  (2,-0.5)*+={\bullet}="2",
  (3,-0.5)*+={\bullet}="3",
  (4,-0.5)*+={\bullet}="4"
  \ar@{=}"0";"1"
  \ar@{=}"1";"2"
  \ar@{=}"3";"4"
  \ar@{-}@`{(2,2)}"0";"4"
  \ar@{-}@`{(2.5,0.5)}"2";"3"
  \endxy \,\circ_2\,
  \xy <0.7cm,0cm>:
  (0,-0.5)*+={\bullet}="0",
  (1,-0.5)*+={\bullet}="1",
  (2,-0.5)*+={\bullet}="2",
  (3,-0.5)*+={\bullet}="3"
  \ar@{:}"0";"1"
  \ar@{:}"1";"2"
  \ar@{:}"2";"3"
  \ar@{.}@`{(1,1)}"0";"2"
  \ar@{.}@`{(2,1)}"1";"3"
  \endxy \,=\, \xy <0.7cm,0cm>:
  (0,-0.5)*+={\bullet}="0",
  (1,-0.5)*+={\bullet}="1",
  (2,-0.5)*+={\bullet}="2",
  (3,-0.5)*+={\bullet}="3",
  (4,-0.5)*+={\bullet}="4",
  (5,-0.5)*+={\bullet}="5",
  (6,-0.5)*+={\bullet}="6",
  (7,-0.5)*+={\bullet}="7"
  \ar@{=}"0";"1"
  \ar@{:}"1";"2"
  \ar@{:}"2";"3"
  \ar@{:}"3";"4"
  \ar@{=}"4";"5"
  \ar@{=}"6";"7"
  \ar@{-}@`{(3.5,2)}"0";"7"
  \ar@{.}@`{(2,1)}"1";"3"
  \ar@{.}@`{(3,1)}"2";"4"
  \ar@{-}@`{(5.5,0.5)}"5";"6"
  \endxy $$
  $$ \xy <0.7cm,0cm>:
  (0,-0.5)*+={\bullet}="0",
  (1,-0.5)*+={\bullet}="1",
  (2,-0.5)*+={\bullet}="2",
  (3,-0.5)*+={\bullet}="3"
  \ar@{=}"0";"1"
  \ar@{=}"2";"3"
  \ar@{-}@`{(1,1)}"0";"2"
  \ar@{-}@`{(2,1)}"1";"3"
  \endxy \,\circ_{(2,4)}\,
  \xy <0.7cm,0cm>:
  (0,-0.5)*+={\bullet}="0",
  (1,-0.5)*+={\bullet}="1",
  (2,-0.5)*+={\bullet}="2",
  (3,-0.5)*+={\bullet}="3"
  \ar@{:}"0";"1"
  \ar@{:}"2";"3"
  \ar@{.}@`{(1.5,1.5)}"0";"3"
  \ar@{.}@`{(1.5,0.5)}"1";"2"
  \endxy \,=\, \xy <0.7cm,0cm>:
  (0,-0.5)*+={\bullet}="0",
  (1,-0.5)*+={\bullet}="1",
  (2,-0.5)*+={\bullet}="2",
  (3,-0.5)*+={\bullet}="3",
  (4,-0.5)*+={\bullet}="4",
  (5,-0.5)*+={\bullet}="5"
  \ar@{=}"0";"1"
  \ar@{:}"1";"2"
  \ar@{=}"3";"4"
  \ar@{:}"4";"5"
  \ar@{-}@`{(1.5,1.5)}"0";"3"
  \ar@{.}@`{(3,2)}"1";"5"
  \ar@{.}@`{(3,1)}"2";"4"
  \endxy $$
  \end{center}
  \caption{Compositions along a unpaired base and a pairing}
\end{figure}


By definition the compositions preserve the type of the first argument,
i.e. if it is a 0/1-structure, then the resulting structure again has type 0/1.

Obviously composition with an identity $\id_k$ has no effect and composition with
the empty structures $\empt_k$ deletes the strucutral element along which it 
is composed.


Since the composition along an unpaired base or a pairing does not affect the 
remaining structural elements, we may compose simultanously along all of
them. This is denoted by $\sigma \circ (\tau_1, \dots , \tau_l)$,
where $\tau_i$ is composed with $\sigma$ along the $i$-th structural element,
ordered by $<_{\sigma}$.




As already mentioned, the compositions provide means to build larger structures
from smaller ones. To get an efficient description, one needs to use a set
of basic structures, called {\em generators}, and to restrict to
{\em decomposable structures}, i.e. those which are compositions of
generators, reducing the search space. 

\begin{definition}[{{\bf Decomposable structure}}]
  Let $\Gamma$ be a finite set of 0- and 1-structures. Its elements are called {\em generators}.
  A structure $\sigma$ is {\em $\Gamma$-decomposable} if
  \begin{enumerate}
    \item it is an identity ($\id_0$ and $\id_1$),
    \item or there exists a generator $\tau\in\Gamma$ and $\Gamma$-decom\-pos\-able
    structures $\sigma_i$ for $1\leq i \leq l$, such that $\sigma = \tau \circ(\sigma_1,
    \dots, \sigma_l)$. 
  \end{enumerate}
\end{definition}

\begin{figure}
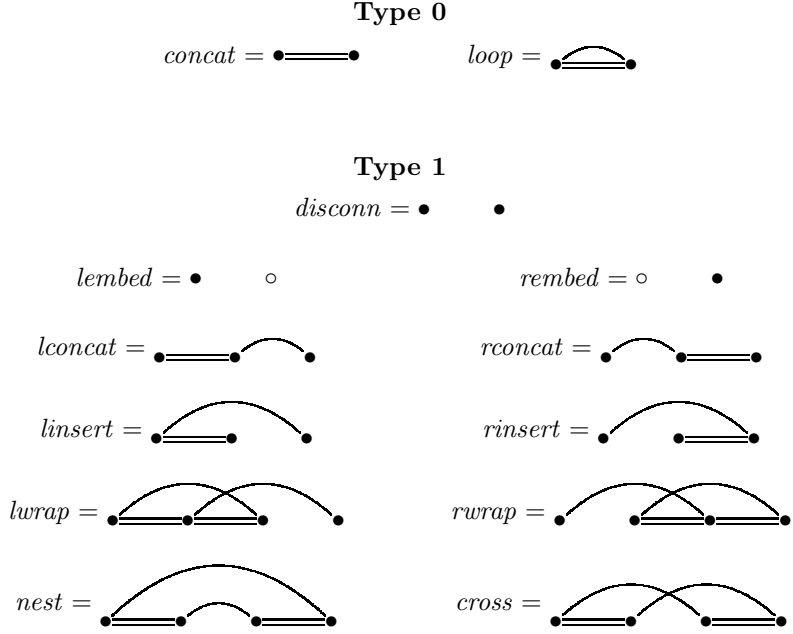

\begin{center}
{\bfseries Type 0}
$$
\begin{array}{c@{\qquad\qquad}c}
\concat = \xy<1cm,0cm>:
  (0,0)*+={\bullet}="0",
  (1,0)*+={\bullet}="1"
  \ar@{=}"0";"1"
\endxy 
&
\lp = \xy <0cm,-0.125cm>;<1cm,-0.125cm>:
  (0,0)*+={\bullet}="0",
  (1,0)*+={\bullet}="2"
  \ar@{=}"0";"2"
  \ar@{-}@`{(0.5,0.5)}"0";"2"
\endxy  
\end{array}
$$
\vspace*{5mm}

{\bfseries Type 1}
$$
\begin{array}{c@{\qquad\qquad}c}
  \multicolumn{2}{c}{
    \disconn = \xy <1cm,0cm>:
    (0,0)*+={\bullet}="0",
    (1,0)*+={\bullet}="1"
    \endxy  
  }
  \\[5mm]
  \lembed = \xy <1cm,0cm>:
  (0,0)*+={\bullet}="0",
  (1,0)*+={\circ}="1"
  \endxy  
  &
  \rembed = \xy <1cm,0cm>:
  (0,0)*+={\circ}="0",
  (1,0)*+={\bullet}="1"
  \endxy  
  \\[5mm]
  \lconcat =
  \xy <0cm,-0.125cm>;<1cm,-0.125cm>:
  (0,0)*+={\bullet}="0",
  (1,0)*+={\bullet}="1",
  (2,0)*+={\bullet}="2"
  \ar@{=}"0";"1"
  \ar@{-}@`{(1.5,0.5)}"1";"2"
  \endxy 
  &
  \rconcat = 
\xy <0cm,-0.125cm>;<1cm,-0.125cm>:
(0,0)*+={\bullet}="0",
  (1,0)*+={\bullet}="1",
  (2,0)*+={\bullet}="2"
  \ar@{=}"1";"2"
  \ar@{-}@`{(0.5,0.5)}"0";"1"
\endxy 
\\[5mm]

\linsert =
\xy <0cm,-0.125cm>;<1cm,-0.125cm>:
  (0,0)*+={\bullet}="0",
  (1,0)*+={\bullet}="1",
  (2,0)*+={\bullet}="2"
  \ar@{=}"0";"1"
  \ar@{-}@`{(1,1)}"0";"2"
\endxy 
&
\rinsert = 
\xy <0cm,-0.125cm>;<1cm,-0.125cm>:
  (0,0)*+={\bullet}="0",
  (1,0)*+={\bullet}="1",
  (2,0)*+={\bullet}="2"
  \ar@{=}"1";"2"
  \ar@{-}@`{(1,1)}"0";"2"
\endxy 
\\[5mm]

\lwrap =
\xy <0cm,-0.125cm>;<1cm,-0.125cm>:
  (0,0)*+={\bullet}="0",
  (1,0)*+={\bullet}="1",
  (2,0)*+={\bullet}="2",
  (3,0)*+={\bullet}="3"
  \ar@{=}"0";"1"
  \ar@{=}"1";"2"
  \ar@{-}@`{(1,1)}"0";"2"
  \ar@{-}@`{(2,1)}"1";"3"
\endxy 
&
\rwrap = 
\xy <0cm,-0.125cm>;<1cm,-0.125cm>:
  (0,0)*+={\bullet}="0",
  (1,0)*+={\bullet}="1",
  (2,0)*+={\bullet}="2",
  (3,0)*+={\bullet}="3"
  \ar@{=}"1";"2"
  \ar@{=}"2";"3"
  \ar@{-}@`{(1,1)}"0";"2"
  \ar@{-}@`{(2,1)}"1";"3"
\endxy 
\\[5mm]

\stack = \xy <0cm,-0.25cm>;<1cm,-0.25cm>:
  (0,0)*+={\bullet}="0",
  (1,0)*+={\bullet}="1",
  (2,0)*+={\bullet}="2",
  (3,0)*+={\bullet}="3"
  \ar@{=}"0";"1"
  \ar@{=}"2";"3"
  \ar@{-}@`{(1.5,0.5)}"1";"2"
  \ar@{-}@`{(1.5,1.5)}"0";"3"
\endxy
&
\inter = \xy <0cm,-0.25cm>;<1cm,-0.25cm>:
  (0,0)*+={\bullet}="0",
  (1,0)*+={\bullet}="1",
  (2,0)*+={\bullet}="2",
  (3,0)*+={\bullet}="3"
  \ar@{=}"0";"1"
  \ar@{=}"2";"3"
  \ar@{-}@`{(1,1)}"0";"2"
  \ar@{-}@`{(2,1)}"1";"3"
\endxy 
\end{array}
$$
\end{center}
\caption{The generators\label{figgenerators}}
\end{figure}

For biological purposes we suggest the generators given in Fig. \ref{figgenerators}.
In \cite{bauer04studienarbeit} experiments confirmed that all pseudoknots in
PseudoBase \cite{pseudobase} are decomposable with respect to these generators.
This is a strong indication that indecomposable structures are biologically
less relevant.

The generator $\disconn$ allows the construction of disconnected 1-structures, 
consisting of two 0-structures. 

\begin{figure}[b]
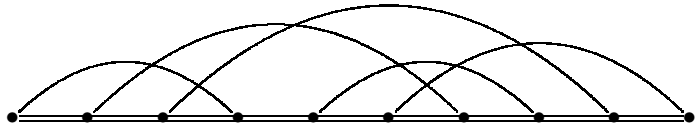

\begin{center}
$
\xy <1cm,0cm>:
  (0,0)*+={\bullet}="0",
  (1,0)*+={\bullet}="1",
  (2,0)*+={\bullet}="2",
  (3,0)*+={\bullet}="3",
  (4,0)*+={\bullet}="4",
  (5,0)*+={\bullet}="5",
  (6,0)*+={\bullet}="6",
  (7,0)*+={\bullet}="7",
  (8,0)*+={\bullet}="8",
  (9,0)*+={\bullet}="9"
  \ar@{=}"0";"9"
  \ar@{-}@`{(1.5,1.5)}"0";"3"
  \ar@{-}@`{(3.5,2.5)}"1";"6"
  \ar@{-}@`{(5,3)}"2";"8"
  \ar@{-}@`{(5.5,1.5)}"4";"7"
  \ar@{-}@`{(7,2)}"5";"9"
\endxy
$
\end{center}
\caption{\label{fignondecomposable}A non-decomposable structure}
\end{figure}

Observe, that all nested structures can be described as compositions of
$\concat$, $\lp$, $\lconcat$, $\rconcat$ and $\stack$. Nonetheless, not all
structures can be build from the generators in $\G$ (eg. the structure shown
in Fig. \ref{fignondecomposable}).


\section{Alignments}

One way to measure the similarity between two folded sequences is the
calculation of the edit distance, i.e. the minimal number (or score) of allowed
(usually reversible) operations needed to construct one sequence from the
other. In the unstructured case this distance is precisely the score of an
{\em alignment} of the two sequences (\cite{gusfield97book}).
This also holds for folded sequences, as long as one restricts to
certain edit operations, which are:
\begin{enumerate}
\item {\em base-replacement} or {\em base-mismatch}, replacing the character
  at one unpaired base with a different one
\item {\em base-deletions}, removing an unpaired base
\item {\em base-insertions}, adding an unpaired base
\item {\em pair-replacement}, replacing the characters at the ends of a pairing
  and changing at least one of them
\item {\em pair-deletion}, removing both ends of a pairing
\item {\em pair-insertion}, adding a pairing
\end{enumerate}

For the representation of inserted and deleted characters we use {\em blanks}
$\blank$. For an arbitrary alphabet $\Sigma$ let $\bar\Sigma$ be the disjoint
union of $\Sigma$ with the blank $\{\blank\}$ (Graphically we represent blanks
by empty circles, while non-blanks (i.e. characters) are represented by discs).

For an arbitrary $\bar\Sigma$-sequence $s$ the $\Sigma$-sequence $\pi(s)$
is obtained by removing all occurrences of the blank from $s$. In the same way
the folded $\Sigma$-sequence $\pi(\sigma,s)$ may be constructed from a folded
$\bar\Sigma$-sequence $(\sigma,s)$, i.e. all bases of $\sigma$ associated with
a blank are removed. If at least one base of a pairing is associated to a
blank, the whole pairing is deleted.

\begin{figure}
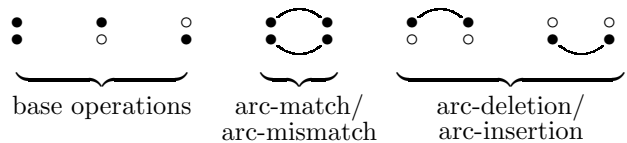

\begin{center}
$$
\xy<0.75cm,0cm>:
  (0,0)*+={\bullet}="0",
  (1.5,0)*+={\bullet}="1",
  (3,0)*+={\blank}="2",
  (4.5,0)*+={\bullet}="3",
  (5.5,0)*+={\bullet}="4",
  (7,0)*+={\bullet}="5",
  (8,0)*+={\bullet}="6",
  (9.5,0)*+={\blank}="7",
  (10.5,0)*+={\blank}="8",
  (0,-0.3)*+={\bullet}="a0",
  (1.5,-0.3)*+={\blank}="a1",
  (3,-0.3)*+={\bullet}="a2",
  (4.5,-0.3)*+={\bullet}="a3",
  (5.5,-0.3)*+={\bullet}="a4",
  (7,-0.3)*+={\blank}="a5",
  (8,-0.3)*+={\blank}="a6",
  (9.5,-0.3)*+={\bullet}="a7",
  (10.5,-0.3)*+={\bullet}="a8",
  (1.5,-1)*=<2.25cm,0mm>\frm{_\}},
  (1.5,-1.5)*{\text{base operations}},
  (5,-1)*=<1cm,0mm>\frm{_\}},
  (5,-1.5)*{\text{arc-match/}},
  (5,-1.9)*{\text{arc-mismatch}},
  (8.75,-1)*=<3cm,0mm>\frm{_\}},
  (8.75,-1.5)*{\text{arc-deletion/}},
  (8.75,-1.9)*{\text{arc-insertion}},

   \ar@{-}@`{(5,0.5)}"3";"4"
   \ar@{-}@`{(7.5,0.5)}"5";"6"

   \ar@{-}@`{(5,-0.8)}"a3";"a4"
   \ar@{-}@`{(10,-0.8)}"a7";"a8"
\endxy 
$$
\end{center}
\caption{The graphical representation of the edit operations\label{graphedit}}
\end{figure}

A (structural) alignment is the description of two folded sequences obtained
from each other by a sequence of basic edit operations, without remembering
them in detail. But in fact a possible edit sequence between both structures
may easily be constructed from an alignment, and vice versa.

\begin{definition}[{{\bf Alignment}}]
  Let $(\sigma_k,s_k) = ((n_k,P_k),s_k), k=1,2$, be two folded
  sequences of same type. A {\em (structural) alignment}
  $(\sigma,t_1,t_2)$ between $(\sigma_1,s_1)$ and $(\sigma_2,s_2)$
  is a structure $\sigma$ of same type, together with two $\bar\Sigma$-sequences
  $t_1, t_2$, such that 
  \begin{itemize}
    \item $(\sigma_k,s_k) = \pi(\sigma,t_k)$ for $k=1,2$ and
    \item $(s_k[i],s_k[j]) \in \Sigma\times\Sigma\sqcup\{(\blank,\blank)\}$
      for $(i,j)\in\sigma$ and $k=1,2$.
  \end{itemize}
\end{definition}

\begin{figure}[b]
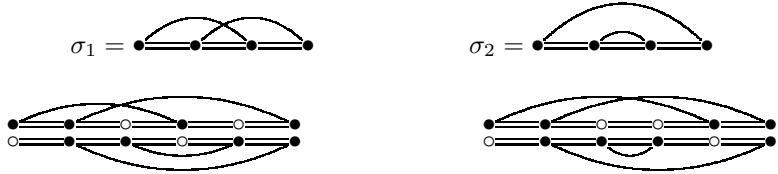

\begin{center}
$
\hfill
\sigma_1 = \xy<0.75cm,0cm>:
  (0,0)*+={\bullet}="0",
  (1,0)*+={\bullet}="1",
  (2,0)*+={\bullet}="2",
  (3,0)*+={\bullet}="3"
  \ar@{-}@`{(1,1)}"0";"2"
  \ar@{-}@`{(2,1)}"1";"3"
  \ar@{=}"0";"1"
  \ar@{=}"1";"2"
  \ar@{=}"2";"3"
\endxy
\hfill
\sigma_2 = \xy<0.75cm,0cm>:
  (0,0)*+={\bullet}="0",
  (1,0)*+={\bullet}="1",
  (2,0)*+={\bullet}="2",
  (3,0)*+={\bullet}="3"
  \ar@{-}@`{(1.5,1.5)}"0";"3"
  \ar@{-}@`{(1.5,0.5)}"1";"2"
  \ar@{=}"0";"1"
  \ar@{=}"1";"2"
  \ar@{=}"2";"3"
\endxy
\hfill
$

\vspace{5mm}

$
\hfill
\xy<0.75cm,0cm>:
  (0,0)*+={\bullet}="0",
  (1,0)*+={\bullet}="1",
  (2,0)*+={\blank}="2",
  (3,0)*+={\bullet}="3",
  (4,0)*+={\blank}="4",
  (5,0)*+={\bullet}="5",
  (0,-0.3)*+={\blank}="a0",
  (1,-0.3)*+={\bullet}="a1",
  (2,-0.3)*+={\bullet}="a2",
  (3,-0.3)*+={\blank}="a3",
  (4,-0.3)*+={\bullet}="a4",
  (5,-0.3)*+={\bullet}="a5"
  \ar@{-}@`{(1.5,0.75)}"0";"3"
  \ar@{-}@`{(3,1)}"1";"5"
  \ar@{-}@`{(3,-1.3)}"a1";"a5"
  \ar@{-}@`{(3,-0.8)}"a2";"a4"
  \ar@{=}"0";"1"
  \ar@{=}"1";"2"
  \ar@{=}"2";"3"
  \ar@{=}"3";"4"
  \ar@{=}"4";"5"
  \ar@{=}"a0";"a1"
  \ar@{=}"a1";"a2"
  \ar@{=}"a2";"a3"
  \ar@{=}"a3";"a4"
  \ar@{=}"a4";"a5"
\endxy
\hfill\hfill
\xy<0.75cm,0cm>:
  (0,0)*+={\bullet}="0",
  (1,0)*+={\bullet}="1",
  (2,0)*+={\blank}="2",
  (3,0)*+={\blank}="3",
  (4,0)*+={\bullet}="4",
  (5,0)*+={\bullet}="5",
  (0,-0.3)*+={\blank}="a0",
  (1,-0.3)*+={\bullet}="a1",
  (2,-0.3)*+={\bullet}="a2",
  (3,-0.3)*+={\bullet}="a3",
  (4,-0.3)*+={\blank}="a4",
  (5,-0.3)*+={\bullet}="a5"
  \ar@{-}@`{(2,1)}"0";"4"
  \ar@{-}@`{(3,1)}"1";"5"
  \ar@{-}@`{(3,-1.3)}"a1";"a5"
  \ar@{-}@`{(2.5,-0.8)}"a2";"a3"
  \ar@{=}"0";"1"
  \ar@{=}"1";"2"
  \ar@{=}"2";"3"
  \ar@{=}"3";"4"
  \ar@{=}"4";"5"
  \ar@{=}"a0";"a1"
  \ar@{=}"a1";"a2"
  \ar@{=}"a2";"a3"
  \ar@{=}"a3";"a4"
  \ar@{=}"a4";"a5"
\endxy \hfill $
\end{center}
  \caption{\label{figalignments}Two structural alignments of two folded sequences}
\end{figure}

More intuitively, an alignment consists of a structure $\sigma$, such that
the structures $\sigma_1$ and $\sigma_2$ can be obtained by removing bases and
pairings. The bases and pairings which have to be removed are indicated by the
position of the blanks in the sequences. The second condition ensures that
either both bases in a pairing are associated to blanks, or none. Two examples of
alignments are shown in Fig. \ref{figalignments}.

Instead of counting the minimal number of edit operations needed for the
transformation of one sequence into another, we assign a non-negative score
$S(\sigma,t_1,t_2)$ to an alignment and try to minimize it. The
score is the sum of scores of the structural elements of $\sigma$, depending
on both, the corresponding bases of the aligned structures, and the underlying
structural elements. For each unpaired base of the alignment we use scores
$$ S \astack{x}{y}\qquad S \astack{x}{\blank} \qquad S \astack{\blank}{y}$$
where $x$ and $y$ are arbitrary characters. For pairings we have 
scores of the following forms.
$$
   S\pstack{x_1}{}{x_2}{y_1}{}{y_2} \qquad 
   S\pstack{x_1}{}{x_2}{\blank}{}{\blank} \qquad
   S\pstack{\blank}{}{\blank}{y_1}{}{y_2}
$$
The {\em score} of an alignment $(\sigma,t_1,t_2)$ is defined by
$$ S(\sigma,t_1,t_2) := \sum_{i\in\sigma}
S\astack{t_1[i]}{t_2[i]} + \sum_{(i,j)\in \sigma} S
\pstack{t_1[i]}{\,}{t_1[j]}{t_2[i]}{\,}{t_2[j]}.
$$
In addition to the scores being non-negative, we require 
$$ S\astack{\circ}{\circ} = S\astack{x}{x} = 0 = S \pstack{x}{}{y}{x}{}{y} =S \pstack{\circ}{}{\circ}{\circ}{}{\circ}. $$

The score of a minimum alignment of two folded sequences is
$$ S\left((\sigma_1,s_1),(\sigma_2,s_2)\right) := \min \left\{
    S(\sigma,t_1,t_2) \mid \pi(\sigma,t_k) = (\sigma_k,s_k), k=1,2  \right\}.
$$

\subsection{Semi-decomposable Alignments}

As shown in \cite{zhang02similarity} the general problem of finding the score of
a minimum alignment is NP-complete. But we are going to describe an algorithm
calculating the exact minimum alignment of a decomposable folded sequence and 
an arbitrary one, using the fact that an arbitrary alignment of these is {\em
semi-decomposable}, as defined below.
 
In the following $\Gamma$ is a finite set of generators including those in Fig. 
\ref{figgenerators}. The notions of semi- and  decomposability will always refer
to this set. 

Let $P$ be a set of pairings in $\sigma$, then $\sigma\setmins P$ is the structure 
obtained by removing all pairings (including their bases) in $P$. If $(\sigma,s)$
is a folded sequence we define $(\sigma,s)\setmins P := (\sigma\setmins P,
s\setmins P)$, where $s\setmins P$ is constructed from $s$ by removing all
letters associated to the bases of pairings in $P$. 

As we will see an alignment of a decomposable folded sequence with an arbitrary
one is semi-decomposable in the following sense. This observation and Lemma 
\ref{lemsplitting} give us a possibility to calculate the minimal alignment by
decomposition.

\begin{definition}[{{\bf Semi-decomposable}}]
	An alignment $(\sigma,t_1,t_2)$ of two folded sequences $(\sigma_1,s_1)$ and $(\sigma_2,s_2)$
	is called {\em semi-decomposable} if there exists a set $P$ of pairings in $\sigma$, such
	that 
	\begin{enumerate}
	\item $(t_k[i],t_k[j]) = (\circ,\circ)$ for one $k\in\{1,2\}$ and each $(i,j)\in P$,
	\item and the structure $\sigma\setmins P$ is decomposable.
	\end{enumerate}
\end{definition}

For an arbitrary folded $0$-sequence $(\sigma,s)$ of length $n$ and $1\leq i < j \leq n$ 
let $\sigma[i, j]$ be the structure obtained from $\sigma$ by removing all unpaired bases
outside the interval $[i,j]$ {\bf and} all pairings with at least one end outside of it.
Furthermore we define $(\sigma,s)[i,j] := (\sigma[i,j],\bar s)$, where $\bar s$ is obtained
from $s$ by deletion of the letters assigned to the deleted bases. Similar the
$1$-structure $(\sigma,s)[i_1,j_1;i_2,j_2]$ is defined for $i_1\leq j_1<i_2
\leq j_2$. 


Let $\tau$ be an arbitrary structure with $m$ bases and $\sigma$ one of same type with $n$ 
bases. A {\em $\tau$-splitting} of $\sigma$ is a partition of the interval $[1,n]$ into $m$
subintervals $I^1, \dots, I^m$ with $I^l=[i^l,j^l]$, which respect the gaps if $\tau$ and 
$\sigma$ have type 1, i.e.
\begin{itemize}
\item $i^1 = 1$, $ j^k+1 = i^{k+1}$ for $1 \leq k < m$ and $j^m = n$,
\item $j^l = k$ if $\sigma$ if the gap in $\tau$ is between
	$l$ and $l+1$ and the gap in $\sigma$ between $k$ and $k+1$.
\end{itemize}
A pairing $(i,j)$ of $\sigma$ is called {\em incompatible} with the splitting,
if $i\in I^{i'}$ and $j\in J^{j'}$ with $i`\neq j`$ and $(i',j')$ is not a pairing
in $\tau$, i.e. the two bases of the pairing lie in two different intervals of
the splitting, which aren't paired in $\tau$. A $\tau$-splitting of $\sigma$ is called
{\em proper} if the induced splitting of $\tau\setmins P$ contains no empty interval, where
$P$ is the set of incompatible pairings. In other words a $\tau$-splitting is proper if and 
only if each interval contains at least one unpaired base or one end of a compatible pairing.

The notion of proper splittings allows an equivalent description of semi-decomposable
structures.
\begin{lemma}
	\label{lemsplitting}
	An alignment $(\sigma,t_1,t_2)$ of $(\sigma_1,s_1)$ and $(\sigma_2,s_2)$ is
	semi-decomposable, if and only if either $\sigma$ is an identity, a generator, 
	or if
	\begin{enumerate}
	\item there exists a generator $\tau$ with $m$ bases and $\iota$ structural elements,
	\item proper $\tau$-splittings $I_k^1, \dots I_k^m$ of $\sigma_k$ for $k=1,2$,
	\item for each structural element $\chi$ of $\tau$ exists a semi-decomposable 
		aligment $(\sigma^{\chi},t_1^{\chi},t_2^{\chi})$ of $(\sigma_1,s_1)[I_1^i]$ and 
		$(\sigma_2,s_2)[I_2^i]$ if $\chi = i\in\tau$, or of $(\sigma_1,s_1)[I_1^i;I_1^j]$ 
		and $(\sigma_2,s_2)[I_2^i;I_2^j]$ if $\chi = (i,j)\in\tau$,
	\item and $\sigma_k\setmins P_k = \tau\circ(\pi(\sigma^1,t_k), \dots, \pi(\sigma^{\iota},t_k))$
	for $k=1,2$, where $P_k$ is the set of {\em incompatible} pairings of $\sigma_k$.
	\end{enumerate}
\end{lemma}
\begin{proof}
	If $\sigma$ is an identity or a generator the lemma obviously holds.
	
	Suppose that the alignment $(\sigma,t_1,t_2)$ is semi-decomposable. If $\sigma$ is only an 
	unpaired base, then it obviously is a decomposition and $P`=\emptyset$. The same holds if 
	$\sigma$ has only a single pairing.
	Now assume that $P$ is the set of pairings defined in the definition of semi-decomposability, ie.
	$\sigma\setmins P = \tau \circ (\sigma^1, \dots, \sigma^m)$ for a generator $\tau$ and decomposable 
	structures $\sigma^1, \dots, \sigma^m$. Then one can chose a $\tau$-splitting of $\sigma$, such that
	the induced splitting of $\tau\setmins P$ is the one given by the decomposition (simply add the bases
	of the deleted pairings to appropriate intervals). Since the generators do not allow the deletion
	of bases or pairings, this splitting is proper and some of the pairings in $P$ are compatible
	with it and some aren't. Let $P'\subseteq P$ be the subset of the latter ones. The
	compatible pairings can be added to the $\sigma^i$ resulting in structures $\hat\sigma^1, \dots, 
	\hat\sigma^m$. By induction these induce semi-decomposable subalignments, because they contain
	less structural elements than $\sigma$ and removal of the added pairings results in decomposable 
	structures. Furthermore we have $\sigma\setmins P' = \tau \circ (\hat\sigma^1, \dots, \hat\sigma^m)$.
	
	Now assume that there exists a generator $\tau$ and a proper $\tau$-splitting as stated in the lemma.
	Since the subalignments are semi-decomposable and contain less structural elements than $\sigma$, there
	exists a set $P_{\chi}$ of pairings in $\sigma^{\chi}$ for each structural element $\chi\in\tau$, such that
	$\sigma^{\chi}\setmins P_{\chi}$ is decomposable and each pairing in $P_{\chi}$ is assigned to blanks 
	in at least one sequence. If $P$ is the set of pairings incompatible to the splitting, this leads to
	$$ \sigma \setmins P\cup \bigcup_{\chi\in\tau} P_{\chi} = \tau \circ \left( \sigma^1\setmins P_1 ,\dots
	, \sigma^{\iota}\setmins P_{\iota} \right), $$
	proving the semi-decomposability of $(\sigma,t_1,t_2)$.
\end{proof}

\begin{corollary}
  \mylabel{corsemidecomp}
  Any alignment $(\sigma,t_1,t_2)$ of a decomposable folded sequence $(\sigma_1,s_1)$
  and an arbitrary folded sequence $(\sigma_2,s_2)$ is semi-decomposable.  
\end{corollary}
\begin{proof}
	Choose $P'$ as the set of pairings $(i,j)$ in $\sigma$ matched by blanks, ie. $(t_2[i], t_2[j]) = (\circ,\circ)$.
	Then $\sigma\setmins P'$ is exactly $\sigma_1$ with additional unpaired bases. Since unpaired bases may be added
	at any position by composition with generators, $\sigma\setmins P'$ is decomposable.
\end{proof}


As a consequence of Corollary \ref{corsemidecomp}, it is sufficient to find the
minimum semi-de\-compos\-able alignment if at least one of the sequences is
decomposable. An arbitrary semi-decomposable alignment can be constructed 
in the following way.
\begin{enumerate}
\item Choose a generator $\tau$.
\item Choose two proper $\tau$-splittings of the structures.
\item Find all incompatible pairings.
\item Align the subsequences (without incompatible pairings) induced by the splittings.
\end{enumerate}
Then the score of the alignment is the sum of the scores of the subalignments and
the scores of incompatible pairings matched against blanks. This approach leads to the
algorithm described in detail in the following section.

\subsection{The Algorithm}

The minimum alignment of two folded sequences is 
calculated using dynamic programming. We use two arrays, indexed by intervals. For 
two folded $0$-sequences $(\sigma_1,s_1)$ and $(\sigma_2,s_2)$ the value $S_0[I_1;I_2]$
is the score of a minimal alignment of the $0$-structures $(\sigma_1,s_1)[I_1]$ and
$(\sigma_2,s_2)[I_2]$. Similar $S_1[I_1,J_1;I_2,J_2]$ is the score 
of a minimal alignment of the two $1$-sequences $(\sigma_1,s_1)[I_1;J_1]$ and
$(\sigma_2,s_2)[I_2;J_2]$.
The values $R_k[I;J]$ are the sums of all {\em weights} of
pairings $(i,j)$ in $(\sigma_k,s_k)$ such that one end is in $I$ and the other
in $J$, i.e.
$$ R_k[I;J] := \sum_{(i,j)\in \sigma_k, i\in I, j\in J} 
S \pstack{s_k[i]}{}{s_k[j]}{\blank}{}{\blank}. $$ 
$W(\sigma,s)$ is the score of the sequence $(\sigma,s)$ (aligned with
blanks), i.e.
$$ W(\sigma,s) := \sum_{i\in\sigma} S\astack{s[i]}{\blank} +
\sum_{(i,j)\in\sigma} S\pstack{s[i]}{}{s[j]}{\blank}{}{\blank}.$$ 
For shorter notation we write:
$$ W_k[I] := W\left( (\sigma_k,s_k)[I]\right) \text{ and } W_k[I;J] := W
\left( (\sigma_k,s_k)[I;J] \right)$$

The entries of the arrays $S_0$ and $S_1$ can be calculated recursively using scores for
shorter intervals. The recursion stops if all intervals consist of only one base, i.e. 
$I_k = [i_k,i_k]$ 
and $J_k = [j_k,j_k]$. Then we have:
\begin{equation}
	\label{eq01}
  S_0[I_1;I_2] =
  \begin{cases} 
    \min \left( S\astack{s_1[i_1]}{s_2[i_2]}, W_1[I_1] + W_2[I_2] \right)
    & \text{ if $i_1$ and $i_2$ are unpaired} \\[3mm]
    W_1[I_1] + W_2[I_2] & \text{otherwise} 
  \end{cases}
 \end{equation}
\begin{multline}
\label{eq02}
  S_1[I_1,J_1;I_2,J_2] \\
  = \left\{\begin{array}{rlr}
  &\multicolumn{2}{l}{\min\left\{
    S\pstack{s_1[i_1]}{}{s_1[j_1]}{s_2[i_2]}{}{s_2[j_2]} , W_1[I_1;J_1]
  + W_2[I_2;J_2] \right\}} \\
  && (i_1,j_1) \text{ and } (i_2,j_2) \text{ are pairings} \\[2mm] 
  &W_1[I_1;J_1] + W_2[I_2] + W_2[J_2] 
  & \text{only $(i_1,j_1)$ is a pairing} \\[2mm] 
  &W_2[I_2;J_2] + W_1[I_1] + W_1[J_1] 
  & \text{ only $(i_2,j_2)$ is a pairing} \\[2mm] 
  &W_1[I_1] + W_1[J_1] + W_2[I_2] + W_2[J_2] 
  & \text{otherwise} 
 \end{array}\right.
\end{multline}

In general, we have to check every generator $\tau$ of type $0$ and every pair
$ I_k = I_k^1 \dots I_k^m$, $k=1,2$, of proper $\tau$-splittings of $\sigma_k[I_k]$.
The score of this decomposition is
\begin{multline}
 \label{eqs0}
 X_0(\tau,I_k^1, \dots ,I_k^m) := \\
 \sum_{i\in\tau}S_0[I_1^i;I_2^i]
 + \sum_{(i,j)\in\tau}S_1[I_1^i,I_1^j;I_2^i,I_2^j]
 + \sum_{(i,j)\not\in\tau} \left( R_1[I_1^i; I_1^j] + R_2[I_2^i; I_2^j] \right).
\end{multline}
The first two sums are the scores contributed by subalignments induced by the
unpaired bases and pairings in $\tau$. The third sum is the score of incompatible
pairings.

Since the splittings are proper, the intervals $I_k^i$ aren't empty and therefore are
shorter than $I_k$. Hence $X_0(\tau,I_k^1, \dots ,I_k^m)$ can be calculated from the
scores of subaligments of shorter intervals.

The score $S_0[I_1;I_2]$ is the minimum over all generators of type 0 and all
proper splittings. For $\concat$ and $\lp$ this leads to
$$
  S_0[I_1;I_2]
  = \min \begin{cases}
     S_0[I_1^1;I_2^1] + S_0[I_1^2;I_2^2] &  (\tau = \concat) \\
     \qquad + R_1[I_1^1;I_1^2] + R_2[I_2^1;I_2^2] \\[2mm]
     S_1[I_1^1,I_1^2;I_2^1,I_2^2] & (\tau = \lp)
     \end{cases}
$$
where $I_k=I_k^1I_k^2, k=1,2,$ are decompositions of the intervals.

Similar, $S_1$ can be calculated as the minimum score over all decompositions using a
generator $\tau$ of type 1. In general, one obtains the following
score for $S_1[I_1,J_1;I_2,J_2]$ if $\tau$ has non-empty legs with $m$ and $l$ bases, 
$m,l \leq 1$:
\begin{multline}
 \label{eqs1}
 X_1(\tau,I_k^1, \dots ,I_k^{m+l}) :=  \\
 \sum_{i\in\tau}S_0[I_1^i;I_2^i] + \sum_{(i,j)\in\tau}S_1[I_1^i,I_1^j;I_2^i,I_2^j]
  + \sum_{(i,j)\not\in\tau} \left( R_1[I_1^i; I_1^j] + R_2[I_2^i; I_2^j] \right).
\end{multline}
where $I_k = I_k^1\dots I_k^m$, $J_k = J_k^{m+1} \dots J_k^{m+l}$ for $k=1,2$.


Again the score $S_1[I_1,J_1;I_2,J_2]$ is the minimum over all sums for all generators of type 1
and all appropriate splittings.
For some of the generators given in Fig. \ref{figgenerators} this results in the formulas 
given in Tab. \ref{figw1}.


\begin{table}
 \begin{multline*}
 S_1[I_1,J_1;I_2,J_2] = \\
 \min \begin{cases}
     S_0[I_1;I_2] + S_0[J_1;J_2]  & \hfill (\tau = \disconn) \\ \qquad
     + R_1[I_1;J_1] + R_2[I_2;J_2] \\[1.5mm]
%
%
%
     S_0[I_1^1;I_2^1] + S_1[I_1^2,J_1;I_2^2,J_2]
     & \hfill \text{ for } I_1 = I_1^1I_1^2 \text{ and } I_2 = I_2^1I_2^2 \\ \qquad
     + R_1[I_1^1;I_1^2] + R_1[I_1^1;J_1] & \hfill I_k^l \neq I_k 
     \\ \qquad
     + R_2[I_2^1;I_2^2] + R_2[I_2^1;J_2] & \hfill (\tau = \lconcat) \\[1.5mm]
%
%
%
%
     S_1[I_1^1,I_1^3;I_2^1,I_2^3] + S_1[I_1^2,J_1;I_2^2,J_2] 
     & \hfill \text{ for } I_1 = I_1^1I_1^2I_1^3 \text{ and } I_2 = I_2^1I_2^2I_2^3 \\ \qquad
     + R_1[I_1^1;I_1^2] + R_1[I_1^2;I_1^3] & \hfill I_k^2 \neq I_k 
     \\ \qquad
     + R_1[I_1^1;J_1] + R_1[I_1^3;J_1] & \hfill (\tau = \lwrap) \\ \qquad
     + R_2[I_2^1;I_2^2] + R_2[I_2^2;I_2^3] \\ \qquad
     + R_2[I_2^1;J_2] + R_1[I_2^3;J_2] \\[1.5mm]
%
%
     S_1[I_1^1,J_1^2;I_2^1,J_1^2] + S_1[I_1^2,J_1^1;I_2^2,J_1^1]  
     & \hfill \text{ for } I_k=I_k^1I_k^2, J_k=J_k^1J_k^2 
     \\ \qquad
     + R_1[I_1^1;I_1^2] + R_1[I_1^1;J_1^1] & \hfill I^k_l \neq I^k, J^k_l \neq
     J_k 
     \\ \qquad
     + R_1[I_1^2;J_1^2] + R_1[J_1^1;J_1^2] & \hfill (\tau = \stack) \\ \qquad
     + R_2[I_2^1;I_2^2] + R_2[I_2^1;J_2^1] \\ \qquad
     + R_2[I_2^2;J_2^2] + R_2[J_2^1;J_2^2] \\[1.5mm]
     S_1[I_1^1,J_1^1;I_2^1,J_1^1] + S_1[I_1^2,J_1^1;I_2^2,J_1^2]  
     & \hfill \text{ for } I_k=I_k^1I_k^2, J_k=J_k^1J_k^2 
     \\ \qquad
     + R_1[I_1^1;I_1^2] + R_1[I_1^1;J_1^2] & \hfill I^k_l \neq I^k, J^k_l \neq
     J^k 
     \\ \qquad
     + R_1[I_1^2;J_1^1]  + R_1[J_1^1;J_1^2] & \hfill (\tau = \inter) \\ \qquad
     + R_2[I_2^1;I_2^2] + R_2[I_2^1;J_2^2] \\ \qquad
     + R_2[I_2^2;J_2^1] + R_2[J_2;J_2^2]
   \end{cases}
 \end{multline*}
\caption{\label{figw1} $S_1$ for the generators $\disconn$, $\lconcat$, $\lwrap$, $\stack$ and $\inter$} 
\end{table}

\begin{algorithm}
  \caption{Calculation of the score of a minimal strucutral alignment}
   \KwIn{Two folded sequences $(\sigma_k,s_k), k=1,2$ of type 0, one of them has to be decomposable}
   \KwOut{The score of a minimum structural alignment between both}
   \BlankLine
    \Begin{
	$n_1 \leftarrow$ number of bases in $\sigma_1$\;
	$n_2 \leftarrow$ number of bases in $\sigma_2$\;
	Mark all entries $S_0[I_1;I_2]$ and $S_1[I_1,J_1;I_2,J_2]$ as undefined\;
	\Return{$S_0((\sigma_1,s_1),(\sigma_2,s_2),(1,n_1),(1,n_2))$}\;
    }
\end{algorithm}

\begin{function}
  \caption{$S_0$($S_1, S_2, I_1, I_2$)}
  \KwIn{Two folded 0-sequences $S_k = (\sigma_k,s_k)$ and two intervals $I_k$, $k=1,2$}
  \KwOut{The score of a minimum structural aligment of $(\sigma_1,s_1)[I_1]$ and $(\sigma_2,s_2)[I_2]$}
  \BlankLine
  \Begin{
  	\lIf{$S_0[I_1;I_2]$ is defined}{
	  \Return{$S_0[I_1;I_2]$}\;
	}
	\ForAll{generators $\tau$ of type 0}{
		\ForAll{proper $\tau$-splittings $I_k = I^1_k \dots I^m_k$, $m=|\tau|$}{
	      $x \leftarrow X_0(\tau,I^1_k, \dots I^m_k)$ \emph{as given by Eqn. (\ref{eq01}) or (\ref{eqs0})}\;
			\lIf{$x < S_0[I_1;I_2]$}{
		   	$S_0[I_1;I_2] \leftarrow x$\;
			}
	   }
	}
	\Return{$S_0[I_1;I_2]$}\;
  }
\end{function}

\begin{function}
  \caption{$S_1$($S_1, S_2, I_1, J_1 , I_2, J_2$)}
  \KwIn{Two folded 0-sequences $S_k = (\sigma_k,s_k)$ and four intervals $I_k, J_k$, $k=1,2$}
  \KwOut{The score of a minimum structural aligment of $(\sigma_1,s_1)[I_1,J_1]$ and $(\sigma_2,s_2)[I_2, J_2]$}
  \BlankLine
  \Begin{
  	\lIf{$S_0[I_1, J_1;I_2, J_2]$ is defined}{
	  \Return{$S_0[I_1, J_1;I_2, J_2]$}\;
	}
	\ForAll{generators $\tau$ of type 1}{
		\ForAll{proper $\tau$-splittings $I_k = I^1_k \dots I^m_k$, $J_k = I^{m+1}_k \dots I^{m+l}_k$}{
	     $x \leftarrow X_1(\tau,I^1_k, \dots, I^{m+l}_k)$ \emph{ as given by Eqn. (\ref{eq02}) or (\ref{eqs1})}\;
			\lIf{$x < S_1[I_1,J_2;I_2,J_2]$}{
		   	$S_1[I_1,J_1;I_2,J_2] \leftarrow x$\;
			}
	   }
	}
	\Return{$S_1[I_1,J_1;I_2,J_2]$}\;
  }
\end{function}

\begin{theorem}
  Let $\Gamma$ be a set of generators, i.e. a finite set of 0- and 1-structures, including those in
  Fig. \ref{figgenerators}, and $m$ the maximal number of bases in any leg of any generator $\tau\in \Gamma$.
  The score of a minimum alignment of a $\Gamma$-decomposable folded sequence with
  $n_1$ bases, and an arbitrary one with $n_2$ bases, can be calculated in
  $O(n_1^{m+4}n_2^{m+4})$ time and $O(n_1^4n_2^4)$ space.
\end{theorem}
\begin{proof}
There are $O(n_k^2)$ weights $W((\sigma_k,s_k)[I])$, which can be
calculated in $O(n_k)$ each, leading to $O(n_1^3 + n_2^3)$ time and
$O(n_1^2+n_2^2)$ space. Similar the $R_k[I;J]$ need $O(n_1^4+n_2^4)$ space and
$O(n_1^5+n_2^5)$ total time. 

The arrays $S_0$ and $S_1$ have $O(n_1^2 n_2^2)$ and $O(n_1^4n_2^4)$ entries.
We assume that the calculation of $X_0(\tau,\dots)$ and $X_1(\tau,\dots)$ using
equations (\ref{eqs0}) and (\ref{eqs1}) require constant time (this is the case
if the equations for each generator $\tau$ are directly programmed and not 
dynamically evaluated). Furthermore already known scores are stored such that
each score has to be calculated at most once.

Since the $R_k[I;J]$ and the weights $W$ are already known, the time required for
each splitting of the intervals is constant, and the check wether a splitting is proper
takes $O(n_1+n_2)$ time. Hence it is sufficient to count the splittings for each entry. 
For $S_0$ each interval is split into at most $m$ parts, resulting in $O(n_1^{m-1}n_2^{m-1})$ 
cases and $O(n_1^{m+2}n_2^{m+2})$ time. For $S_1$ each interval is splitted into at most
$m$ subintervals, leading to $O(n_1^{m-1}n_2^{m-1})$ cases and $O(n_1^{m+4}n_2^{m+4})$ time
for $S_1$.
\end{proof}

\begin{corollary}
  For the set $\Gamma$ of generators given in Fig. \ref{figgenerators},
  the score of a minimum alignment of a decomposable folded sequence with
  $n_1$ bases, and an arbitrary one with $n_2$ bases, can be calculated in
  $O(n_1^7n_2^7)$ time and $O(n_1^4n_2^4)$ space.
\end{corollary}

\subsection{The Approximation of arbitrary alignments}

Obviously the calculations can be made for two arbitrary folded sequences,
resulting in a minimum semi-decomposable alignment. As we have seen, this
leads to a globally minimum alignment, if at least one of them is
decomposable. But what can be said about the calculated score if neither structure
is decomposable.

Let $S_{\text{decomp}} = S_0[(1,n_1);(1,n_2)]$ be the score calculated by our
algorithm. Then we have the following result regarding the quality of the approximation.
\begin{theorem}
  If there exists a constant $c\geq 1$, such that
  $$ c S\pstack{x}{}{y}{x'}{}{y'} \geq S\pstack{x}{}{y}{\blank}{}{\blank} +
  S\pstack{\blank}{}{\blank}{x'}{}{y'} $$ for all $x,y,x',y' \in \Sigma$ with
  $x\neq x'$ or $y\neq y'$, then 
  $$ S_{\text{decomp}}((\sigma_1,s_1),(\sigma_2,s_2)) \leq c
  S((\sigma_1,s_1),(\sigma_2,s_2))$$
  for all folded sequences $(\sigma_k,s_k), k=1,2$.
\end{theorem}
\begin{proof}
  Let $(\sigma,t_1,t_2)$ be a minimum alignment. We are going to construct a
  semi-decomposable alignment, replacing all mismatched pairings by an
  insertion and a deletion. This procedure allows us to control the 
  increase of the score and leads to the stated inequality, and it ensures that
  the constructed alignment remains semi-decomposable, since its decomposable
  core is constructed from the initial alignment by the deletion of pairings.

  Let $M = \{ (i,j) \in \sigma \mid t_1[i] \neq  t_2[i] \text{ or } t_1[j]
  \neq t_2[j] \text{ and } t_k[i],t_k[j] \neq \blank \}$ be the set of
  mismatched pairings. Then we have: 
  \begin{align*}
    S((\sigma_1,s_1), (\sigma_2,t_2))
    &= S(\sigma,t_1,t_2) \\
    & = \underbrace{\sum_{i\in\sigma} S\astack{t_1[i]}{t_2[i]}}_{:=A} + 
    \sum_{(i,j)\in\sigma} S\pstack{t_1[i]}{}{t_1[j]}{t_2[i]}{}{t_2[j]}  \\
    & = A + \underbrace{\sum_{(i,j)\not\in M}    S\pstack{t_1[i]}{}{t_1[j]}{t_2[i]}{}{t_2[j]}}_{:=B} +
    \sum_{(i,j)\in M} S\pstack{t_1[i]}{}{t_1[j]}{t_2[i]}{}{t_2[j]}  \\
    & \geq A + B
    + \sum_{(i,j)\in M} \frac{1}{c}\left( S\pstack{t_1[i]}{}{t_1[j]}{\blank}{}{\blank} +
      S\pstack{\blank}{}{\blank}{t_2[i]}{}{t_2[j]} \right)  \\ 
    & \geq \frac{1}{c} ( A + B) \\
    &  \qquad
        + \frac{1}{c}\sum_{(i,j)\in M} \left( S\pstack{t_1[i]}{}{t_1[j]}{\blank}{}{\blank} +
      S\pstack{\blank}{}{\blank}{t_2[i]}{}{t_2[j]} \right) \\
    & \geq \frac{1}{c} S_{\text{decomp}}((\sigma_1,s_1),(\sigma_2,s_2)) 
  \end{align*}
\end{proof}

If the scores satisfy
$$ S\pstack{x}{}{y}{x'}{}{y'} \leq S\pstack{x}{}{y}{\blank}{}{\blank} +
  S\pstack{\blank}{}{\blank}{x'}{}{y'} $$ for at least one combination of
  $x,y,x',y'\in\Sigma$ with $x\neq x'$ or $y\neq y'$, then one may simply
  choose: 
$$ c := \max \left\{ \frac{S\pstack{x}{}{y}{\blank}{}{\blank} +
  S\pstack{\blank}{}{\blank}{x'}{}{y'}}{S\pstack{x}{}{y}{x'}{}{y'}} \mid x\neq
  x' \text{ or } y \neq y' \right\} \geq 1.$$
In particular, one obtains:

\begin{corollary}
  $S((\sigma_1,s_1),(\sigma_2,s_2)) =
  S_{\text{decomp}}((\sigma_1,s_1),(\sigma_2,s_2))$ for arbitrary folded
  sequences $(\sigma_k,s_k), k=1,2$, if
  $$S\pstack{x}{}{y}{x'}{}{y'} = S\pstack{x}{}{y}{\blank}{}{\blank} +
  S\pstack{\blank}{}{\blank}{x'}{}{y'}$$ for $x\neq x'$ or $y\neq y'$.
\end{corollary}


\section{Conclusion}

We presented a new formal description of secondary RNA structures
incorporating a wide variety of pseudoknots, i.e. tertiary structures.
Based on this description it is possible to calculate the exact score of a
minimal structural alignment of a decomposable and an arbitrary structure.
For other cases the algorithm provides an approximation of guaranteed quality,
depending on the chosen weights for the underlying edit operations.

For the set of generators given in Fig. \ref{figgenerators} the algorithm 
requires $O(n^{14})$ time and $O(n^8)$ space, where $n$ is the number of 
bases in the longer sequence. Hence, further improvements
are necessary. For example, the structures may be reduced to their underlying
stems, i.e. sequences of nested pairs, decreasing the number of nodes. On the
other hand, it might be possible to restrict to special decompositions of the 
intervals, which would reduce the required space. This would at least lead to
an approximative algorithm producing exact results in many cases. To which
extent and at which quality remains to be analyzed.

On the other hand the set of structures, for which the algorithm produces
exact results, can be extended by the addition of generators. This causes an
increasing runtime. But as long as only a finite number of generators is used,
the algorithm stays polynomial, even though the general problem is NP-complete.
Hence the addition of generators leads to a sequence of polynomial solvable
problems (possibly) ``converging'' to an NP-complete problem. Hence it may be
interesting to examine the gap between decomposable and non-decomposable
structures.



\bibliographystyle{alpha}
\bibliography{../bib/bibliography.bib}

\end{document}